\documentclass[prd,onecolumn,notitlepage,nofootinbib,superscriptaddress,showpacs,showkeys]{revtex4-2}

\usepackage{amsfonts}
\usepackage{graphicx}
\usepackage{dcolumn}
\usepackage{amsmath}
\usepackage{amssymb}
\usepackage[T1]{fontenc}
\usepackage[utf8]{inputenc}
\usepackage{esint}
\usepackage[brazil,english]{babel}
\usepackage[usenames,dvipsnames]{color}
\usepackage{longtable}
\usepackage{ulem}
\usepackage{nicefrac}
\usepackage{bm}
\usepackage{tikz}
\usepackage{braket}
\usepackage{graphicx, nicefrac}
\usepackage{hyperref}
\usepackage{subfloat}

\usepackage{float}

\usetikzlibrary[shapes,arrows]

\usepackage{cancel} 
\usepackage{ulem} 

\begin{document}

\title{Generalized Thermodynamics and Phase Transitions of Vaidya-adS Black Holes}

\author{A. F. Cardona}
\email{afcardonaji@gmail.com}
\author{C. Molina}
\email{cmolina@usp.br}
\affiliation{Universidade de S\~{a}o Paulo, Escola de Artes, Ci\^{e}ncias e Humanidades, Avenida Arlindo Bettio 1000, CEP 03828-000, S\~{a}o Paulo-SP, Brazil}

\begin{abstract}
This work investigates the thermodynamics of non-stationary anti-de Sitter (adS) black holes and their gauge-gravity duals. To model a non-equilibrium process, we employ a time-dependent Vaidya-adS metric to describe the collapse of a null fluid and the subsequent thermalization of the dual field theory. We introduce an effective version of free energy as a quasilocal probe to be used in a dynamical setup. By analyzing how this quantity evolves, we show that a first-order phase transition occurs when the final black-hole mass exceeds the limit established by the Hawking--Page temperature. Furthermore, we demonstrate that the final state only approaches the thermal behavior of the $\mathcal{N} = 4$ supersymmetric Yang--Mills theory for mass states much larger than the adS radius. Finally, we compute the boundary energy--momentum tensor using holographic renormalization and compare the resulting energy density with the strong-coupling limit. This demonstrates that the dual description provides a consistent framework for both the equilibrium and non-equilibrium thermodynamics of strongly coupled gauge theories.
\end{abstract}


\maketitle

\section{Introduction}
\label{section1}

The study of anti-de Sitter (adS) geometries stands as one of the most fruitful and impactful areas in modern theoretical physics.
This is in large part due to their central role in formulating the anti-de Sitter/conformal field theory (adS/CFT) correspondence, which establishes an equivalence between a gravitational theory in a higher-dimensional ``bulk'' adS space and a lower-dimensional conformal field theory (CFT) on its boundary \cite{Maldacena:1997re,Gubser:1998bc,Witten:1998qj}.

The adS/CFT correspondence was initially constructed as a duality between an $\mathcal{N} = 4$ supersymmetric Yang--Mills theory (SYM) with gauge group $SU(N)$ in the large $N$ limit and type IIB supergravity on $\text{adS}_5 \times S^5$. One characteristic of the adS/CFT correspondence is the possibility of studying a strongly-coupled gauge theory, for which a perturbative analysis is not normally possible, by identifying the gravitational dual. This allows one to relate the physical observables associated with the field theory, (e.g., correlation functions or Wilson loops), to their geometric counterparts.

Conversely, one might ask, given an arbitrary asymptotically adS geometry, what information can be obtained about the field theory on the boundary?
The answer to this question is a procedure known as holographic renormalization \cite{balasubramanian1999stress,de2001holographic}, which allows one to calculate  the energy--momentum tensor on the conformal boundary using the metric from the gravitational side.

Even before the formulation of the adS/CFT correspondence, adS spacetimes were a well-known subject of research due to their geometric and thermodynamic properties. Unlike black holes in flat or de Sitter space, which typically evaporate completely, asymptotically anti-de Sitter black holes can achieve stable thermal equilibrium with radiation. This stability arises because the asymptotic structure of the spacetime acts like a gravitational ``box''. Radiation escaping from the black hole is bent back by the spacetime curvature, falling back onto the horizon. The length scale of the cosmological constant establishes a clear distinction between unstable small adS black holes and large stable adS black holes \cite{hawking1983thermodynamics}.
Furthermore, the unique thermal behavior of adS black holes necessitates more complex thermodynamic frameworks for their description. Several studies have sought to refine our understanding of these features \cite{hawking1999rotation,gpp,kastor2009enthalpy,dolan2011pressure,bfm2,fbfm,hemming2007thermodynamics,Campos:2024fpf,xiao2024extended,Campos:2025nnk,Cardona:2026kjc}.

Another remarkable characteristic of adS geometries is the existence of phase transitions between different thermal configurations. The most well-known transition in adS gravity is the Hawking--Page phase transition \cite{hawking1983thermodynamics}, which occurs between a low-temperature phase, where empty adS space coexists with a dilute gas of thermal radiation, and a high-temperature phase dominated by a stable, large adS black hole. Even though much of the discussion of phase transitions of adS black holes involves the description of equilibrium states associated with stationary geometries, the thermalization process itself is dynamic, and it should be expected that deviations from equilibrium are described by non-stationary geometries. 

In this work, we adopt a framework that accounts for the time-dependence of the black-hole mass and horizon. Using a Vaidya--anti-de Sitter (Vaidya-adS) metric as a basis provides a setup to track the evolving black hole.
We introduce an effective version of free energy as a quasilocal probe to be used in a dynamical regime. By analyzing how this quantity evolves, we can study the transition between different thermodynamic states over time. These results are complemented by an investigation of the energy--momentum tensor at the adS timelike boundary. 
The quasilocal diagnostic tool is as an effective method for tracking thermodynamic preference rather than a strict microscopic dynamic transition, contrasting with non-local boundary probes \cite{balasubramanian2011thermalization,Baron:2013eja,hubeny2014holographic,abajo2010holographic,lin2021nonequilibrium}.

The remainder of this paper is organized as follows. 
In Section~\ref{section2}, we review the Vaidya-adS spacetime and its thermodynamic properties.
In Section~\ref{section3}, a specific mass profile is introduced and the time-dependent behavior of the free energy is analyzed. 
Section~\ref{section4} is dedicated to the calculation and analysis of the boundary energy--momentum tensor associated with the Vaidya-adS geometry. 
Concluding remarks are presented in Section~\ref{sec:conclusion}.
In this work, we adopt the metric signature $(-,+,\cdots,+)$ and work in natural units in which $c = \hbar = 1$.

\section{Non-Stationary \text{adS} Spacetime}
\label{section2}

To introduce the time-dependent thermalization of an asymptotically adS black hole, we employ a Vaidya-adS spacetime \cite{vaidya1951gravitational, wang1999generalized,xiang2000entropy}. This geometry can be employed to model the gravitational collapse via an infalling thin-shell profile. 
The ($d+1$)-dimensional Vaidya-adS metric with cosmological constant $\Lambda = - \nicefrac{3}{L^{2}}$ in advanced Eddington--Finkelstein coordinates $(v, r,\theta_{1},\theta_{2},\ldots,\theta_{d-1})$ is as follows:
\begin{equation}\label{adS_metric}
ds^2 = -f(r, v) dv^2 + 2 dr dv + r^2 d\Omega_{d-1}^2 \,,
\end{equation}
where 
\begin{equation}\label{adS_function}
f(r, v) = 1 + \frac{r^2}{L^2} - \frac{\mu(v)}{r^{d-2}} \,.
\end{equation}
The usual radial coordinate is denoted by $r$. The coordinate $v$ represents the advanced time (where surfaces of constant $v$ are incoming null hypersurfaces), and the function $\mu(v)$ is proportional to the time-dependent mass profile $M(v)$:
\begin{equation}
\mu(v) = \frac{16\pi G_N M(v)}{(d-1)\Omega_{d-1}} \,.
\end{equation}
The line element of the unit sphere $S^{d-1}$,
\begin{equation}
d\Omega_{d-1}^{2}=\left(d\theta_{1}\right)^{2}+\sin^{2}\theta_{1}\,\left(d\theta_{2}\right)^{2}+\cdots+\sin^{2}\theta_{1}\cdots\,\sin^{2}\theta_{d-1}\,\left(d\theta_{d-1}\right)^{2} \,,
\label{dOmega2}
\end{equation}
is parametrized by the angular coordinates $\left\{ \theta_{1},\theta_{2},\ldots,\theta_{d-1}\right\}$. The hypervolume of $S^{d-1}$ is denoted by $\Omega_{d-1}$.

The metric presented in Equation~\eqref{adS_function} is sourced by a null dust energy--momentum tensor:
\begin{equation}
T_{\mu\nu} = \rho(v, r) \, k_\mu k_\nu\,, \quad \rho(v, r) = \frac{d-1}{16\pi G_N r^{d-1}} \frac{d\mu(v)}{dv}\,,
\end{equation}
with $k_\mu = -\partial_\mu v$.  
Enforcing the weak energy condition, implemented here as $\rho \ge 0$, results in the monotonic, non-decreasing behavior of the mass profile, such that $\nicefrac{d\mu}{dv} \ge 0$.

When dealing with a non-stationary black-hole geometry, the thermodynamic prescription must consider the time-dependency of the metric. We follow the generalized thermodynamics formalism introduced by Hayward et al. \cite{Hayward:1994,Hayward:1996,Hayward:1998sh}. In this approach, the thermodynamic quantities are associated with the trapping horizon $(r_{\text{H}})$, which is defined locally as the outermost marginally trapped surface, where the expansion of outgoing null geodesics vanishes. 
In spherically symmetric spacetimes, this definition is foliation-independent, making it an appropriate choice for a quasi-local identification of the black-hole horizon.
For the Vaidya-adS metric in Equations~\eqref{adS_metric} and \eqref{adS_function}, this condition is satisfied when the metric function~\eqref{adS_function} vanishes:
\begin{equation}\label{Apparent_horizon}
f(r_{\text{H}}(v), v) = 0 \implies 1 + \frac{r_{\text{H}}^2}{L^2} - \frac{\mu(v)}{r_{\text{H}}^{d-2}} = 0 \,.
\end{equation}

In Hayward's formalism of generalized thermodynamics, the dynamic surface gravity $\kappa(v)$ of the trapping horizon $r_{\text{H}}(v)$ is defined via the gradient of $f(r,v)$:
\begin{equation}\label{dynamic_kappa}
\kappa(v) = \left. \frac{1}{2} \frac{\partial f(r, v)}{\partial r} \right|_{r = r_{\text{H}}(v)} = \frac{d\, r_{\text{H}}(v)^2 + (d-2)L^2}{2 L^2 r_{\text{H}}(v)}\,.
\end{equation}
The dynamic temperature and Bekenstein--Hawking entropy associated with the trapping horizon are given by the following:
\begin{equation}\label{dynamic_temperature}
T_{\text{dyn}} (v) = \frac{\kappa(v)}{2\pi} \,, 
\quad 
S(r_{\text{H}}) = \frac{\Omega_{d-1} r_{\text{H}}^{d-1}}{4 G_N}\,.    
\end{equation}

Black-hole configurations are thermodynamically admissible only for temperatures $T$ satisfying 
\begin{equation}
T_{dyn} \ge T_{\min} = \frac{\sqrt{d(d-2)}}{2\pi L} \, .
\end{equation}
Below this critical threshold, black-hole solutions are absent, and the system resides exclusively in the stable thermal adS phase.

\section{Non-Equilibrium Phase Transitions }
\label{section3}

\subsection{Time-Dependent Free Energy}

In adS gravity, the standard Hawking--Page transition is a first-order phase transition between thermal adS (gas of particles) and a large adS black hole at a critical Hawking temperature: \cite{hawking1983thermodynamics}
\begin{equation}
T_{\text{HP}} = \frac{d-1}{2\pi L} \,.    
\end{equation}

In the generalized thermodynamic formalism, the phase transition  from thermal adS gas to an adS black hole can be modeled as an incoming shell, or as a continuous, spherically symmetric flux of radiation injected at the boundary $r \to \infty$, which propagates inward and triggers the dynamic formation of a black-hole horizon \cite{balasubramanian2011thermalization,Baron:2013eja,hubeny2014holographic,abajo2010holographic,lin2021nonequilibrium}.
To study the non-stationary transition from pure thermal gas to a large black hole above the Hawking--Page temperature $T_{\text{HP}}$, we introduce the following $\mu$-profile, based on a smooth step function:
\begin{equation}\label{mu_profile}
\mu(v) = \frac{\mu_f}{2} \left[ 1 + \tanh\left( \frac{v - v_0}{\Delta v} \right) \right] \,.
\end{equation}
The parameter $v_0$ defines the symmetric midpoint of the thermalization process, and the parameter $\Delta v$ determines the timescale for the horizon formation. 

The physical picture under consideration is summarized as follows.
At the initial state ($v \ll v_0$), we have $\mu(v) \to 0$, corresponding to pure thermal adS. During the thermalization window ($v_0 - \Delta v\le v \le v_0 + \Delta v$), matter/radiation is injected from the boundary into the bulk until a final state $\mu(v) \to \mu_f$ is achieved. If $\mu_f$ is chosen such that the final radius $r_+ > L$ (meaning $T > T_{\text{HP}}$), the system completes a phase transition into a stable large adS black hole.

To ensure that the thermodynamic quantities associated with the dynamical horizon remain well-defined during thermalization, an adiabatic approximation is employed. In this approximation, radiation is injected into the bulk over a period of time much longer than the characteristic timescales of the spacetime.
In the Vaidya-adS setup, the infalling matter moves at the speed of light. The time it takes for a single shell of light to travel from the boundary has an upper bound given by the pure adS limit, $\Delta v = \nicefrac{\pi L}{2}$. 
Therefore, for the mass profile Equation~\eqref{mu_profile}, the adiabatic approximation is valid when the mass profile varies slowly compared to the horizon light-crossing timescale, i.e., when $\Delta v$ is sufficiently large relative to $L$.

The dynamical quasi-local energy and surface gravity are tracked along the evolving trapping horizon. In spherically symmetric spacetimes of the form Equation~\eqref{adS_metric} with a non-zero cosmological constant, the Misner--Sharp quasi-local energy $E_{\text{MS}}(r,v)$ inside a sphere of radius $r$ at time $v$ is defined as follows:
\begin{equation}
g^{\mu\nu} \partial_\mu r \partial_\nu r = f(r,v) 
= 1 - \frac{2 G_N E_{\text{MS}}(r, v)}{r^{d-2}} + \frac{r^2}{L^2}\,.    
\end{equation}
Evaluated directly at the trapping horizon $r = r_{\text{H}}(v)$ [where $f(r_{\text{H}}, v) = 0$], the Misner--Sharp energy is proportional to the mass function $\mu(v)$:
\begin{equation}\label{Misner-Sharp-mass}
E_{\text{MS}}(r_{\text{H}}(v), v) = \frac{(d-1)\Omega_{d-1}}{16\pi G_N} \mu(v)\,.    
\end{equation}

Using the Misner--Sharp mass $E_{\text{MS}}(r_{\text{AH}})$ and the dynamic Bekenstein--Hawking entropy as defined in Equation~\eqref{dynamic_temperature}, we introduce a quasi-local free energy $F(r_{\text{H}}, v)$ as a time-dependent analogue of the usual thermodynamical free energy definition:
\begin{equation}\label{free_energy_def}
F(r_{\text{H}}, v) = E_{\text{MS}}(r_{\text{H}}(v)) - T_{\text{dyn}}(r_{\text{H}}(v)) \, S(r_{\text{H}}(v)) \,.
\end{equation}
It is important to note that the definition given in Equation~\eqref{free_energy_def} is not an equilibrium thermodynamic potential, but rather a quasi-local quantity evaluated along the evolving trapping horizon in the adiabatic approach. As we will demonstrate, this definition will allows us to track the instantaneous thermodynamic preferences between the black hole and thermal adS phases along the imposed mass profile.

In this work, we mainly focus on the $d=4$ case of the Vaidya-adS metric~\eqref{adS_metric}, since the $\text{adS}_5$ metric is a well-known geometry with an established field dual, the $\mathcal{N}=4$ SYM \cite{Maldacena:1997re,Gubser:1998bc}. 
From Equation~\eqref{Apparent_horizon}, the trapping horizon satisfies $r_{H}^4 + r_{H}^{2}L^{2} = \mu\,L^2$, with the solution given by the following:
\begin{equation}\label{radiusSadS-5d}
 r_{H}^2 (v)= \frac{L^2}{2}\left[ -1 + \sqrt{1 + \frac{4\mu(v)}{L^2}} \right] \,.
\end{equation}
For $d=4$, the quasi-local free energy Equation~\eqref{free_energy_def} becomes the following:
\begin{equation}
F(r_{\text{H}}, v) = \frac{\pi \, r_{H}^2(v)}{8G_5}\left[ 1 - \frac{r_{H}^2(v)}{L^2}\right] \,.
\end{equation}

When $r_{\text{H}}(v) = L$, the instantaneous temperature $T_{\text{dyn}}$ reaches the Hawking--Page temperature, $T_{\text{HP}} = \nicefrac{3}{2\pi L}$, marking the formal thermodynamic transition point where the spacetime transitions into the globally stable large black hole phase.

\begin{figure}[H]
\centering
{{\includegraphics[width=.60\linewidth]{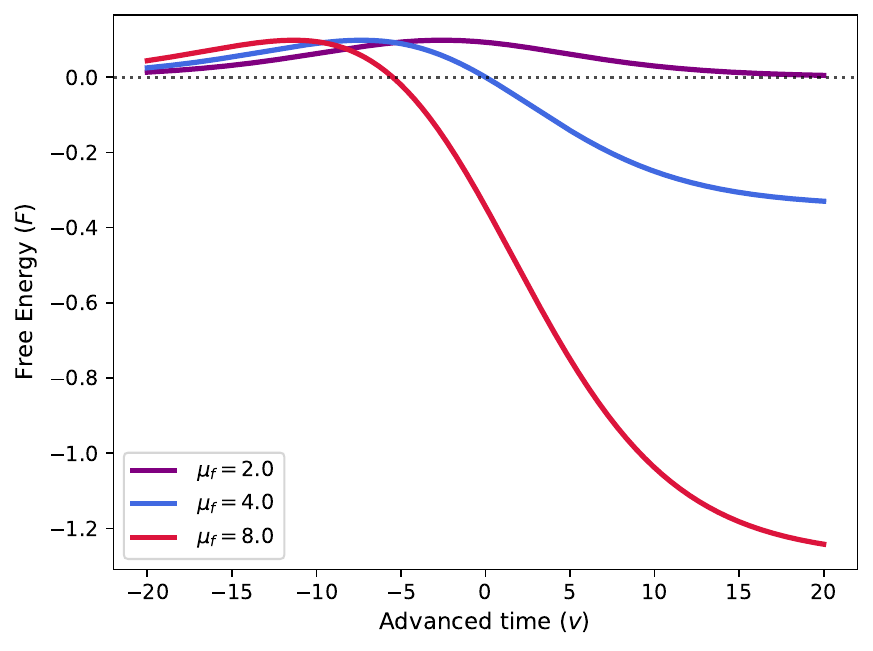}}}
\caption[]{Free energy as a function of advanced time for different values of mass parameter $\mu_f$ with an adS radius $L = 1$. The parameters used for the $\mu$ profile are $v_0 = 0$ and  $\Delta v = 10$, which guarantee the adiabatic approximation.}
\label{fig1}
\end{figure}

\begin{figure}[H]
\centering
{{\includegraphics[width=.60\linewidth]{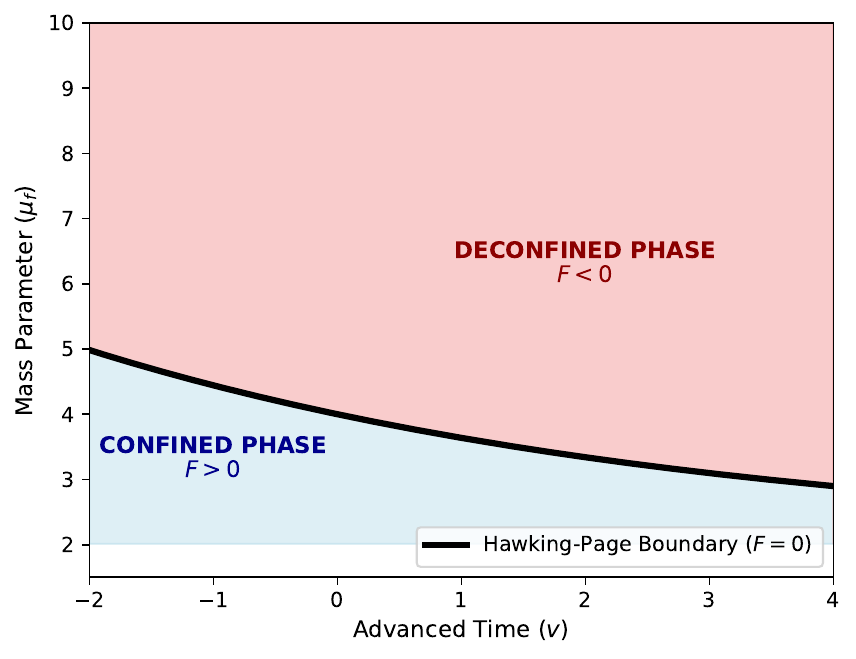}}}
\caption[]{Non-equilibrium phase diagram. The black line indicates the time at which the phase transition occurs. With larger values of injected mass $\mu_f$, the transition triggers at early times.}
\label{fig2}
\end{figure}

Overall, the time-dependent trajectory of the quasi-local free energy~\eqref{free_energy_def} is dictated by the externally imposed mass profile. Setting the free energy of the empty thermal adS background to zero as a baseline, our results imply that a black-hole configuration can only become thermodynamically favored if the free energy becomes negative. Figure~\ref{fig1} shows the evolution of this quantity for different values of mass parameter $\mu_f$.
If the injected mass is not large enough, the free energy will remain positive and the phase transition will not occur. As the mass increases, the transition occurs earlier and the free energy acquires a larger negative value after reaching equilibrium, as illustrated in Figure~\ref{fig2}.

\subsection{Holographic Interpretation in \text{adS/CFT}}

In the context of the adS/CFT correspondence, a $(d+1)$-dimensional Vaidya-adS spacetime is dual to a non-equilibrium thermalization quench in a $d$-dimensional CFT on $S^{d-1} \times \mathbb{R}$ (e.g., $\mathcal{N}=4$ SYM for $\text{adS}_5 \times S^5$).
Under the adS/CFT dictionary for $\text{adS}_5 \times S^5$, the 5D Newton constant relates to the gauge group rank $N$ via the following:
\begin{equation}\label{relation_L_N}
\frac{L^3}{G_N^{(5)}} = \frac{2 N^2}{\pi} \,.    
\end{equation}

The dual CFT exhibits precise, characteristic behaviors regarding the phase transition:
\begin{itemize}
\item A confined phase ($T < T_{\text{HP}}$), with the free energy scaling as $F \sim \mathcal{O}(1)$. The theory consists of color-singlet states (hadrons/glueballs).
\item A deconfined phase ($T > T_{\text{HP}}$), with the free energy scaling as $F \sim \mathcal{O}(N^2)$ and $r_{\text{AH}} > L$. The formation and expansion of the bulk horizon directly manifests as the liberation of the $N^2$ gauge color degrees of freedom, producing a strongly-coupled plasma.
\end{itemize}

In the weak coupling limit of the $\mathcal{N} =4$ SYM, the free energy can be obtained from perturbative calculations as a series of the t'Hooft coupling $\lambda = g^2_{YM} N$ in the large $N$ limit of the gauge group $SU(N)$. 
At leading order, the free energy corresponds to that of non-interacting massless degrees of freedom and is obtained from the sum of one-loop Feynman diagrams of each of the fields of the theory. If the system is at thermal  equilibrium with temperature $T$, the leading free energy is given by the following \cite{kim2000thermodynamics}:
\begin{equation}\label{leading-weak}
F_{0}(T)_{\text{weak}} = -\frac{\pi^2}{6} N^2 V_3 T^4 \,,
\end{equation}
with $V_3$ corresponding to the three-dimensional volume containing the theory.  

At strong coupling $\lambda$, a perturbative calculation of the free energy is not possible. According to the adS/CFT correspondence, the free energy of the $\mathcal{N}=4$ SYM at strong coupling is obtained from the gravitational dual.
Following references \cite{gubser1998coupling,burgess1999free}, the leading contribution to the free energy $F$ for an adS black hole can be obtained from the (euclidean) gravitational action $I$ multiplied by the temperature as $I = \nicefrac{F}{T}$. The leading term in the free energy in the strong-coupling limit is as follows:
\begin{equation}\label{leading-strong}
F_{0}(T)_{\text{strong}} = - \frac{\pi^2}{8}N^2 V_3 T^4 \,.
\end{equation}
By comparing the leading terms in the free energy for the weak and strong coupling limits, Equations~\eqref{leading-weak} and \eqref{leading-strong}, the well-known proportionality relation is obtained: $F_{\text{strong}} = \nicefrac{3}{4} \, F_{\text{weak}}$.

\begin{figure}[H]
\centering
{{\includegraphics[width=.60\linewidth]{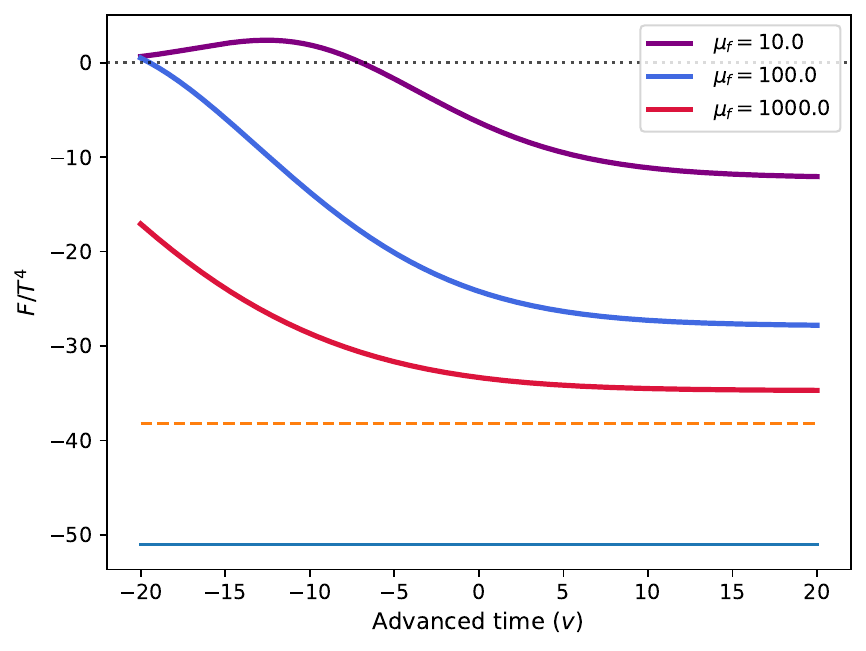}}}
\caption[]{Ratio $\nicefrac{F}{T^4}$ for the Vaidya-$\text{adS}_5$ geometry as a function of advanced time. The dashed line corresponds to the constant $\nicefrac{-\pi^5 L^2}{8G_5}$, with $G_5 = 1$ and $L=1$. The bottom line corresponds to the weak coupling ratio 
$\nicefrac{\pi^5 L^2}{6G_5}$, with $G_5 = 1$ and $L=1$. The parameters used for the $\mu$ profile are $v_0 = 0$ and  $\Delta v = 10$.
}
\label{fig3}
\end{figure}

The calculations for CFT free energy in Equations~\eqref{leading-weak} and \eqref{leading-strong} assume a static background and, thus, thermal equilibrium.
In a non-equilibrium transition, the free energy is no longer a static thermodynamical variable. In particular, it deviates from the $F \sim T^4$ behavior associated to massless non-interactive degrees of freedom. We compare the time-dependent behavior of the free energy in Equation~\eqref{free_energy_def} for the Vaidya-$\text{adS}_5$ geometry with the free energy in the strong coupling limit presented in Equation~\eqref{leading-strong}, or more specifically, with the ratio 
$\nicefrac{F}{T^4} \, $:
\begin{equation}\label{Fstrong_ratio}
\frac{F_{0}(T)_{\text{strong}}}{T^4} = -\frac{\pi^2}{8}N^2 V_3 = -\frac{\pi^5 L^5}{8 G_5} \,,
\end{equation}
where $V_3 = 2\pi^2 L^3$ corresponds to the conformal volume with radius $L$ and Equation~\eqref{relation_L_N} was used to write $N$ in terms of $L$.
In Figure~\ref{fig3}, the time evolution of the ratio $\nicefrac{F}{T^4}$ shows that by increasing orders of magnitude of the mass parameter $\mu(v)$, the behavior of the Vaidya $\text{adS}_5$ free energy approaches in a better fashion the strong coupling result, as indicated by Equation~\eqref{Fstrong_ratio}. 

The adiabatic approach to the strong-coupling plateau, as shown in Figure~\ref{fig3}, is consistent with the behavior observed in holographic quenches with a finite quench duration. For slow-enough mass profiles, the system follows the instantaneous thermal state adiabatically. However, rapid quenches produce non-adiabatic features, such as delayed thermalization and spectral weight transfer \cite{Banerjee:2016ray}.

It is helpful to place our quasi-local free energy probe within the context of the broader program of holographic thermalization. The adS-Vaidya background used in this work is a standard model for the gravitational dual of a conformal field theory, in which the formation of a bulk horizon corresponds to the boundary theory's approach to thermal equilibrium \cite{balasubramanian2011thermalization,Baron:2013eja,hubeny2014holographic,abajo2010holographic,lin2021nonequilibrium}. 
A lesson from the literature is that local gauge-invariant observables may not be sufficient to resolve the thermalization process because their expectation values are determined by the near-boundary behavior of the metric, which remains fixed during the collapse. Non-local probes, such as two-point correlation functions, Wilson loops, and entanglement entropy, are dual to geometric quantities that extend deep into the bulk. Therefore, they can track horizon formation.

The dynamic free energy introduced in Equation~\eqref{free_energy_def} belongs to a different class of probes. Rather than resolving the real-time evolution of specific non-local observables, it provides a quasi-local thermodynamic characterization along the trapping horizon. 
This tool measures which state is currently favored, the black hole or the thermal adS.
In this sense, our approach bridges the Hayward quasi-local formalism \cite{Hayward:1994,Hayward:1996,Hayward:1998sh} and the holographic thermalization scenario. While non-local probes \cite{balasubramanian2011thermalization,Baron:2013eja,hubeny2014holographic,abajo2010holographic,lin2021nonequilibrium} reveal the detailed causal structure and top-down thermalization pattern of the boundary theory, the free energy profile offers a thermodynamic-level diagnostic of when and if the system crosses the Hawking--Page boundary. These two perspectives are therefore complementary, and the free-energy criterion established here is a necessary (though not sufficient) condition for completing the thermalization process.

\section{Boundary Energy--Momentum Tensor}
\label{section4}

Let us complement the development from the previous section with an analysis of the energy--momentum tensor at the boundary of the adS spacetime.
Following the adS/CFT dictionary, the boundary energy-momentum tensor of an asymptotically adS geometry is dual to the vacuum expectation value of the energy-momentum tensor of the CFT \cite{balasubramanian1999stress,de2001holographic}.
By utilizing the prescription of holographic renormalization, the exact non-equilibrium expectation value of the boundary field theory's energy-momentum tensor $\langle T_{\mu\nu} \rangle$ is extracted. 

Any asymptotically adS spacetime with ($d-1$) dimensions can be transformed near the boundary into the following metric in the Fefferman--Graham (FG) coordinates $(z, x^i)$ \cite{bautier2000fefferman}: 
\begin{equation}\label{metric_FG}
 ds^2 = \frac{L^2}{z^2} \left[ \gamma_{ij}(z,x)dx^{i}dx^{j} + dz^2 \right] \,,
\end{equation}
where $z \to 0$ corresponds to the asymptotic boundary and $x^i$ are the $d$-dimensional boundary coordinates ($i, j = 0, \dots, d-1$) \cite{bilic2022thermodynamics}. Near $z=0$, the spatial metric $\gamma_{ij}(z,x)$ expands in powers of z:
\begin{equation}
\gamma_{ij}(z,x) = g^{(0)}_{ij}(x) + z^2\, g^{(2)}_{ij}(x) + \dots + z^{d} g^{(d)}_{ij}(x) + \dots \, ,
\end{equation}
where $g^{(0)}_{ij}(x)$ is the metric on the boundary on which the CFT lives and $g^{(d)}_{ij}(x)$ is the normalizable mode, which contains the dynamical data [such as mass $M(v)$].

To find the explicit series expansion of the metric in Fefferman--Graham coordinates, we must map both the radial coordinate $r$ and advanced time $v$ to the FG radial coordinate $z$ and FG boundary time $t$. 
We assume that the old coordinates can be written as a series expansion of the new coordinates, starting with their standard asymptotic (large $r$) definitions:
\begin{eqnarray}
r(t, z) &=& \frac{L}{z} + r_1(t) z + r_2(t) z^2 + r_3(t) z^3 + r_4(t) z^4 + \dots\ \, , \label{}
\\
v(t, z) &=& t + v_1(t) z + v_2(t) z^2 + v_3(t) z^3 + v_4(t) z^4 + \dots
\, . \label{}
\end{eqnarray}

After matching coefficients with the near boundary behavior ($r \to \infty, z \to 0$) of the metric Equation~\eqref{metric_FG}, explicit expressions are obtained: 
\begin{eqnarray}
r(t,z) &=& \frac{L}{z}\left[ 1-\frac{z^{2}}{4L^{2}} + \frac{\mu (t)}{8L^{4}}z^{4} - \frac{L\dot{\mu}(t)}{24L^{4}}z^{5}+\mathcal{O}(z^{6})\right] \,, \label{radial_mapping}\\
v(t,z) &=& t - \frac{z^{2}}{2L} - \frac{z^{4}}{24L^{3}}+\frac{\mu (t)}{24L^{3}}z^{4}+\mathcal{O}(z^{6}) \label{time_mapping}\,,
\end{eqnarray}
where $\dot{\mu}(t) \equiv \nicefrac{d\mu}{dt}$ denotes the derivative with respect to the boundary time $t$. Substituting Equations~\eqref{radial_mapping} and \eqref{time_mapping} into the metric~\eqref{metric_FG} yields an expansion for each metric component as a power series in $z$. For the time-time component, we have 
\begin{equation}
g_{tt}(t,z) = -\left[1+\frac{1}{2}\left(\frac{z}{L}\right)^{2} + \left(\frac{1}{16}-\frac{3\mu (t)}{4L^{2}}\right)\left(\frac{z}{L}\right)^{4}+\mathcal{O}(z^{5})\right] \,,
\end{equation}
while for the angular three-sphere component, we have the following:
\begin{equation}
g_{\Omega_3}(t,z)=1-\frac{1}{2}\left(\frac{z}{L}\right)^{2}
+ \left[ \frac{1}{16}+\frac{\mu (t)}{4L^{2}}\right] \left(\frac{z}{L}\right)^{4}+\mathcal{O}(z^{5}) \,.
\end{equation}

Since  the boundary geometry is spherically symmetric and isotropic, the spatial cross-terms completely vanish identically ($g_{t\theta} = g_{t\psi} = g_{t\phi} = 0$). From the zeroth-order terms of the expansions, the boundary metric corresponds to the standard Einstein static universe, described by the line element $ds_{\text{boundary}}^{2}=-dt^{2}+L^{2}d\Omega _{3}^{2}$.

From the metric tensor expressed in Fefferman--Graham coordinates, the expected value of the renormalized boundary energy--momentum tensor is \cite{de2001holographic,bautier2000fefferman,bilic2022thermodynamics,apostolopoulos2009cosmology}:
\begin{equation}\label{BoundaryEM}
\langle T_{ij} \rangle =
\frac{d\, L^{d-1}}{16 \pi G_{N}^{(d+1)}}
\left[g_{ij}^{d}(x) + X_{ij}^{(d)}(x)\right] \,,
\end{equation}
where $ X_{ij}^{(d)}(x)$ is equal to zero for $d$ odd. For the case of interest $d = 4$, the boundary energy--momentum tensor assumes the following form:
\begin{equation}
X_{ij}^{(4)}(x) = - \frac{1}{2}g_{i}^{(2)k}g_{kj}^{(2)} 
+ \frac{1}{4}g^{(2)}_{ij}\text{Tr}(g^{(2)})
- \frac{1}{8}g^{(0)}_{ij}\left[ \left(\text{Tr}g^{(2)} \right)^{2} 
-  g_{m}^{(2)n}g_{n}^{(2)m}\right] \,,
\end{equation}
\begin{equation}
\langle T_{ij}\rangle = \frac{L^{3}}{4\pi G_{5}}
\left\{ g^{(4)}_{ij} - \frac{1}{2}g_{i}^{(2)k}g_{kj}^{(2)} 
+ \frac{1}{4}g^{(2)}_{ij}\text{Tr}(g^{(2)})
- \frac{1}{8}g^{(0)}_{ij}\left[ \left( \text{Tr}g^{(2)} \right)^{2} 
-  g_{m}^{(2)n}g_{n}^{(2)m}\right] \right\} \,.   
\end{equation}

After replacing Equations~\eqref{radial_mapping} and \eqref{time_mapping} into the expression for the boundary energy--momentum tensor Equation~\eqref{BoundaryEM}, pure geometric curvature terms cancel out identically, leaving behind a strictly time-dependent energy--momentum tensor of a fluid:
\begin{equation}\label{boundary_EM_tensor}
\langle T_{ij}(t)\rangle dx^{i}dx^{j}=\frac{\mu (t)}{16\pi G_{5}L}\left(3dt^{2}+L^{2}d\Omega _{3}^{2}\right)\, .  
\end{equation}
From Equation~\eqref{boundary_EM_tensor}, the time-dependent energy density and pressure are given by
\begin{equation}\label{boundary_EM_components}
\rho(t) = \langle T^t_{\phantom{t}t} \rangle = \frac{3\mu(t)}{16\pi G_5 L^5} \,, \quad P(t) = \langle T^\theta_{\phantom{\theta}\theta} \rangle = \frac{\mu(t)}{16\pi G_5 L^5} \,.
\end{equation}

\begin{figure}[H]
\centering
{{\includegraphics[width=.60\linewidth]{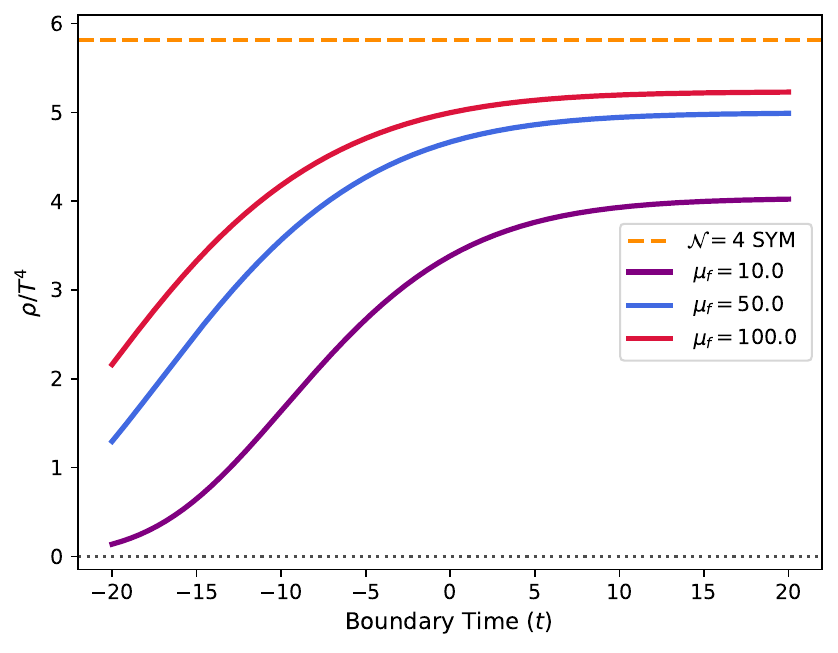}}}
\caption[]{Ratio $\nicefrac{\rho(t)}{T^4}$ as a function of the boundary time for different values of final mass parameter $\mu_f$. The dashed line corresponds to the strongly coupled limit of the gauge theory.}
\label{fig4}
\end{figure}

In Figure~\ref{fig4}, we compare the component $ \langle {T^t}_{t} \rangle$ of the energy--momentum tensor~\eqref{boundary_EM_tensor} with the energy density expected from the strong coupling limit of the gauge theory
\begin{equation}\label{density_strong_coupl}
\rho_{SC} = \frac{3}{8}\pi^2 N^2 T^4 \,.   
\end{equation}
More specifically, we compare the time evolution of the ratio 
$\nicefrac{\rho(t)}{T^4} \,$, with $\rho(t)$ being the time-dependent energy density in Equation~\eqref{boundary_EM_components}, with the corresponding constant obtained from the strong coupling limit in Equation~\eqref{density_strong_coupl}:
\begin{equation}
\frac{\rho_{SC}}{T^4} = \frac{3}{8}\pi^2 N^2 = \frac{3\pi^3L^3}{16{G_5}} \,.   
\end{equation}

We can observe that at early times, small adS black holes are thermodynamically unstable and do not correspond to equilibrium states in the boundary CFT. After the phase transition, the black hole eventually settles to a stable temperature. For increasing values of the final mass parameter, the boundary energy density approaches the exact behavior of the strongly coupled energy density Equation~\eqref{density_strong_coupl}.

\section{Final Remarks}
\label{sec:conclusion}

In this work, we have examined the thermalization dynamics and phase structure of adS black holes by tracking the time evolution of relevant thermodynamic quantities. Exploiting the adS/CFT correspondence, we have compared the thermodynamic properties of a time-dependent adS$_5\times S^5$ geometry with those of the dual $\mathcal{N}=4$ supersymmetric Yang--Mills theory. 
Our analysis was built upon the Vaidya-adS metric, which provides a natural setting for modeling the gravitational collapse of a null fluid and the subsequent thermalization of the dual field theory. 

By adopting the Hayward's generalized thermodynamics formalism, we introduced a dynamical notion of free energy that serves as a quasi-local characterization tool associated with the time-dependent trapping horizon. This approach enabled us to extend some notions from the standard equilibrium thermodynamics description to a genuinely dynamical context.
While non-local boundary probes track the top-down propagation of entanglement entropy from the boundary into the bulk \cite{balasubramanian2011thermalization,Baron:2013eja,hubeny2014holographic,abajo2010holographic,lin2021nonequilibrium}, the trapping horizon free energy tracks the bottom-up, direct formation of the thermal state at the horizon scale. This provides a direct, quasi-local thermodynamic indicator that shows when the bulk black hole becomes thermodynamically competitive relative to the thermal adS background.

Our results show that larger injected masses lead to final equilibrium states that more accurately reproduce the strongly coupled field-theory behavior, while also triggering the Hawking--Page transition earlier.
This establishes an important quantitative criterion: the holographic dual description of non-equilibrium thermalization becomes increasingly accurate when the final black-hole mass and temperature substantially exceed the curvature scale set by $L$.
Through holographic renormalization, the Misner--Sharp mass and the quasi-local free energy are associated with the boundary energy density and the expectation value of the energy--momentum tensor. Thus, the horizon quantities can be understood as proxies for the bulk's thermalization process in the dual-field theory.

We expect the proposed tools to be reliable when the final mass and temperature substantially exceed the curvature scale set by the size of the black-hole horizon. However, we must note the limitations of the method. For instance, near a phase transition where the system is far from equilibrium, the sign and magnitude of the dynamical free energy should be interpreted with caution. While our approach identifies the onset of the thermodynamically favored black-hole phase, it does not establish a dynamical first-order transition on its own.

The presented results in this work have implications for the adS/CFT correspondence as a tool to study real-world strongly coupled systems, such as the quark-gluon plasma produced in heavy-ion collisions.  Our analysis suggests that the Vaidya-adS model, while conceptually simple, captures the essential features of this non-equilibrium dynamics, provided that the final energy scale is sufficiently high.  Our results suggest that future studies of non-equilibrium thermodynamics in adS/CFT should focus on regimes where the injected energy far exceeds the curvature scale. This ensures a clear distinction between the black hole and adS length scales.
It would also be interesting to explore whether the quasi-local free-energy criterion developed here can be extended to non-equilibrium steady states with external fields, where novel non-equilibrium phase transitions have been identified  using the adS/CFT correspondence \cite{Nakamura:2012ae}.

\end{document}